\documentclass[10pt,twocolumn,showpacs,longbibliography,amsmath,amssymb,osajnl,floatfix,superscriptaddress]{revtex4-1}

\usepackage{graphicx}
\usepackage{dcolumn}
\usepackage{bm}
\usepackage{color}
\usepackage{txfonts}
\usepackage{microtype}
\usepackage{siunitx}
\usepackage{url}
\usepackage{epstopdf}
\usepackage{indentfirst}
\usepackage[colorlinks, linkcolor=black,anchorcolor=black,citecolor=black,filecolor=black,urlcolor=blue,bookmarksopen=true]{hyperref}
\usepackage{textcomp}
\usepackage{mathrsfs}
\usepackage{braket}
\usepackage[caption=false,font=normalsize,labelfont=sf,textfont=sf]{subfig}

\begin{document}

\title{Compact Efficient Polarizers for Relativistic Electron Beams}

\author{Kun Xue}
\thanks{These authors contributed equally to this work.}
\affiliation{Key Laboratory for Nonequilibrium Synthesis and Modulation of Condensed Matter (MOE), Shaanxi Province Key Laboratory of Quantum Information and Quantum Optoelectronic Devices, School of Physics, Xi'an Jiaotong University, Xi'an 710049, China}

\author{Yue Cao}
\thanks{These authors contributed equally to this work.}
\affiliation{Department of Physics, National University of Defense Technology, Changsha 410073, China}

\author{Feng Wan}	
\affiliation{Key Laboratory for Nonequilibrium Synthesis and Modulation of Condensed Matter (MOE), Shaanxi Province Key Laboratory of Quantum Information and Quantum Optoelectronic Devices, School of Physics, Xi'an Jiaotong University, Xi'an 710049, China}

\author{Zhong-Peng Li}
\affiliation{Key Laboratory for Nonequilibrium Synthesis and Modulation of Condensed Matter (MOE), Shaanxi Province Key Laboratory of Quantum Information and Quantum Optoelectronic Devices, School of Physics, Xi'an Jiaotong University, Xi'an 710049, China}

\author{Qian Zhao}
\affiliation{Key Laboratory for Nonequilibrium Synthesis and Modulation of Condensed Matter (MOE), Shaanxi Province Key Laboratory of Quantum Information and Quantum Optoelectronic Devices, School of Physics, Xi'an Jiaotong University, Xi'an 710049, China}

\author{Si-Man Liu}
\affiliation{Department of Physics, National University of Defense Technology, Changsha 410073, China}

\author{Xin-Yu Liu}
\affiliation{Department of Physics, National University of Defense Technology, Changsha 410073, China}

\author{Li-Xiang Hu}
\affiliation{Department of Physics, National University of Defense Technology, Changsha 410073, China}

\author{Yong-Tao Zhao}
\affiliation{Key Laboratory for Nonequilibrium Synthesis and Modulation of Condensed Matter (MOE), Shaanxi Province Key Laboratory of Quantum Information and Quantum Optoelectronic Devices, School of Physics, Xi'an Jiaotong University, Xi'an 710049, China}

\author{Zhong-Feng Xu}
\affiliation{Key Laboratory for Nonequilibrium Synthesis and Modulation of Condensed Matter (MOE), Shaanxi Province Key Laboratory of Quantum Information and Quantum Optoelectronic Devices, School of Physics, Xi'an Jiaotong University, Xi'an 710049, China}

\author{Yousef I. Salamin}
\affiliation{Department of Physics, American University of Sharjah, POB 26666, Sharjah, United Arab Emirates}

\author{Tong-Pu Yu}
\email{tongpu@nudt.edu.cn}
\affiliation{Department of Physics, National University of Defense Technology, Changsha 410073, China}

\author{Jian-Xing Li}
\email{jianxing@xjtu.edu.cn}
\affiliation{Key Laboratory for Nonequilibrium Synthesis and Modulation of Condensed Matter (MOE), Shaanxi Province Key Laboratory of Quantum Information and Quantum Optoelectronic Devices, School of Physics, Xi'an Jiaotong University, Xi'an 710049, China}

\date{\today}

\begin{abstract}

Relativistic spin-polarized electron beams are important for fundamental research and the industry, but their generation currently requires conventional accelerators or ultrastrong laser facilities, limiting their accessibility and broad applications. Here, we put forward a novel method for constructing a compact efficient ``polarizer'' that achieves direct ultrafast conversion of relativistic dense electron beams into polarized ones, based on the beam ``self-polarization'' mechanism via simple beam-target interactions. In this scheme, as the electron beam grazes through the polarizer (a double-layer solid target), it ionizes the target and excites an asymmetric plasma field due to the plasma backflows. This field then reacts on the beam itself, triggering  spontaneous radiative polarization and reflection of the beam, and ultimately yielding a dense polarized electron beam. Moreover, the double-layer target setup induces a plasma bubble that focuses the polarized beam and reshapes its polarization distribution. Our method is robust with respect to the beam and target parameters, and opens a new avenue for relativistic beam polarization with compact accessible devices, which would facilitate their broad applications and the development of related experiments, such as in strong-field QED studies, and polarized electron-positron and electron-ion colliders.

\end{abstract}

\maketitle

Spin-polarized electron beams of relativistic energies find important applications in nuclear physics, high energy physics and in the search for new physics beyond the standard model~\cite{glashausser1979nuclear,subashiev1999spin,moortgat2008polarized,remington2006experimental,Jefferson2018precision}, e.g., probing nuclear spin structure~\cite{abe1995precision,maaas2004measurement,alexakhin2007the}, producing polarized gamma rays and positrons~\cite{olsen1959photon,martin2012polarization,abbott2016production,li2020polarized,li2020production}, and studying symmetry violation~\cite{souder1990measurement,anthony2004observation,heckel2006new,heckel2008preferred}. 
Two main methods are currently employed for generating polarized electron beams in experiments.
The first is a two-step process in which the polarized electrons are extracted from  a photocathode~\cite{pierce1976photoemission} or spin filters~\cite{batelaan1999optically, dellweg2017spin}, and then accelerated to relativistic energies in a conventional radio-frequency accelerator~\cite{mane2005spin}.
The second utilizes spontaneous buildup of radiative polarization via the Sokolov-Ternov effect  in high-energy electron storage rings~\cite{sokolov1964on}. 
Both methods rely on conventional accelerators,  which can be quite large, costly, and not readily accessible to normal users~\cite{mane2005spin}.

The rapid progress in modern ultrashort ultraintense laser techniques, achieving peak intensities of $10^{21}-10^{23}$~W/cm$^2$~\cite{kawanaka2016conceptual,Edwin2018,Danson2019,Yoon2021,ELI-beamlines}, has enabled new compact particle acceleration techniques~\cite{esarey2009physics,hooker2013developments,donald2003relativistic,malka1997experimental,gahn1999multi,li2006observation}, and stimulated research into strong-field quantum electrodynamics (QED) phenomena~\cite{yousef2006relativistic,piazza2012extremely,gonoskov2022charged,fedotov2023advances,yu2024bright}, including radiation-reaction effects~\cite{thomas2012strong,cole2018experimental,poder2018experimetal,wistisen2018experimental}, nonlinear Compton scattering (NCS)~\cite{mackenroth2013nonlinear,li2015attosecond,zhao2019ultra}, nonlinear Breit-Wheeler electron-positron pair production~\cite{ridgers2012dense,piazza2016nonlinear,zhu2016dense,xie2017electron,vranic2018mult,tang2021pulse,zhao2022all}, and related polarization effects~\cite{ivanov2004complete,ivanov2005complete,seipt2018theory,king2020nonlinear,li2020production,li2020polarized,wan2019ultra,chen2019polarized,wan2020high,song2022dense,xue2023generation}.
Two laser-driven methods for generating polarized electron beams have been proposed~\cite{buscher2020generation,sun2020production}.
In the first, prepolarized plasma electrons are injected into a plasma wakefield accelerator~\cite{wen2019polarized,wu2019polarizedNJP,nie2021in,nie2022highly,bohlen2023colliding,gong2023spin,sun2024generation,liu2022trapping}, requiring a high-quality prepolarized plasma and careful avoidance of depolarization during injection~\cite{barth2013spin,sofikitis2018ultrahigh,wen2019polarized,fan2022control,thomas2020scaling}.
The second method is radiative polarization of high-energy electrons in an ultraintense asymmetric laser field, such as a bichromatic or elliptically polarized pulse~\cite{seipt2019ultrafast,song2019spin,li2019ultra}, a dual-laser-generated standing wave field~\cite{sorbo2017spin,sorbo2018electron}, or a laser-driven standing wave on the surface of an overdense plasma~\cite{shen2024highlypolarizedenergeticelectrons}.
However, this method has several limitations, including a low average degree of polarization (less than 10\%), the restricted intensity of asymmetric laser pulses due to damage threshold of the optical devices~\cite{tien1999short}, and the need for precise spatiotemporal synchronization between electron beams and laser pulses.
Especially, these laser-based proposals require ultrastrong laser facilities, placing a limit on the otherwise broad number of applications in which polarized beams are perceived to be used. Thus, generation of ultrafast polarization relativistic electron beams with compact and accessible devices remains a great challenge.

\begin{figure}[t]	 
	\centering
	\includegraphics[width=1.0\linewidth]{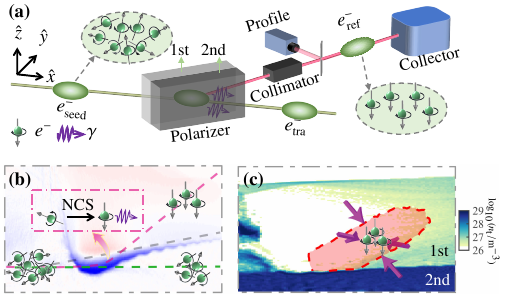}
	\caption{Schematic illustration of the polarizer. (a) A seed electron beam $e^{-}_{\rm seed}$ propagating along $+\hat{{x}}$ interacts with the polarizer, subsequently splitting into two beams: a polarized reflected beam $e^{-}_{\rm ref}$ and an unpolarized transmitted beam $e^{-}_{\rm tra}$. The polarizer consists of two layers [the first (1st) and second (2nd) layers], positioned at a tilt angle $\theta_{\rm t}$ relative to the $xz$ plane; the gray arrows on the electrons indicate their spin direction. (b) Spin-resolved splitting process in the $xy$ plane. The gray-, red- and green-dashed lines indicate the initial interface between the two target layers, and typical trajectories of $e^{-}_{\rm ref}$ and $e^{-}_{\rm tra}$, respectively. These distinct trajectories and polarizations arise from the dependence of their dynamics and polarization on energy loss, as detailed in Fig.~\ref{fig3}. The heat map represents the asymmetric plasma field, which is mainly oriented along $-\hat{y}$ (blue) and $+\hat{y}$ (red). (c) Focusing process in the $xy$ plane. The red arrows represent the focusing force exerted by the plasma ``bubble"; the red area represents the spatial distribution of $e^{-}_{\rm ref}$; the heat map represents the electron density of the target $n_{\mathrm{t}}$.}
	\label{fig1}
\end{figure}

\begin{figure}[t]	
	\centering
	\includegraphics[width=1.0\linewidth]{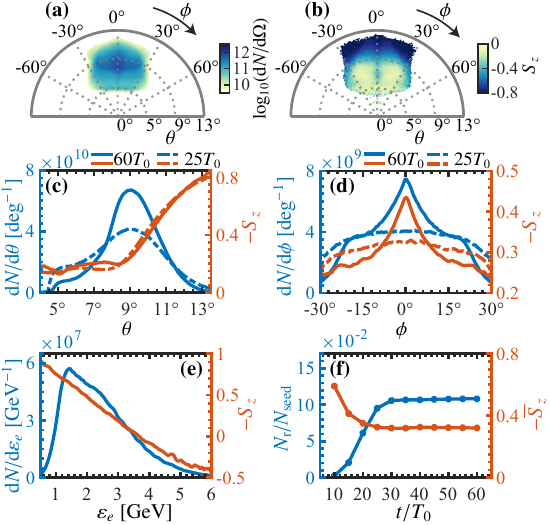}	
	\caption{(a) Angle-resolved electron distribution $\log_{10}({\rm{d}}N_+/{\rm{d}} \Omega)$ and (b) polarization $S_z$ with respect to the polar angle $\theta$ and the azimuthal angle $\phi$, respectively. Here ${\rm d} \Omega$ = $\sin\theta{\rm d}\theta{\rm d}\phi$, $\theta$ = $0^{\circ}$ in $+\hat{{x}}$, and $\phi$ = $0^{\circ}$ in $+\hat{{y}}$. (c) Angle-resolved electron polarization ${S}_z$ (blue) and distribution d$N_{e}$/d$\theta$ (red) vs $\theta$. (d) Angle-resolved polarization ${S}_z$ (blue) and distribution d$N_{e}$/d$\phi$ (red) vs $\phi$. In (c) and (d), the solid and dash-dotted lines represent the results at simulation times $t = 60T_0$ and $25T_0$, respectively. (e) Energy-resolved distribution d$N_{e}$/d$\varepsilon_e$ (blue) and  polarization ${S_z}$ (red) of the electrons at the peak cone angle $\alpha$ vs the electron energy $\varepsilon_e$. (f) Evolution in time of the conversion rate $N_{\rm r}$/$N_{\rm seed}$ (blue) and average polarization ${\overline{S}_z}$ (red) of the reflected electron beam.}
	\label{fig2}
\end{figure}

\begin{figure}[t]	
	\centering
	\includegraphics[width=1.0\linewidth]{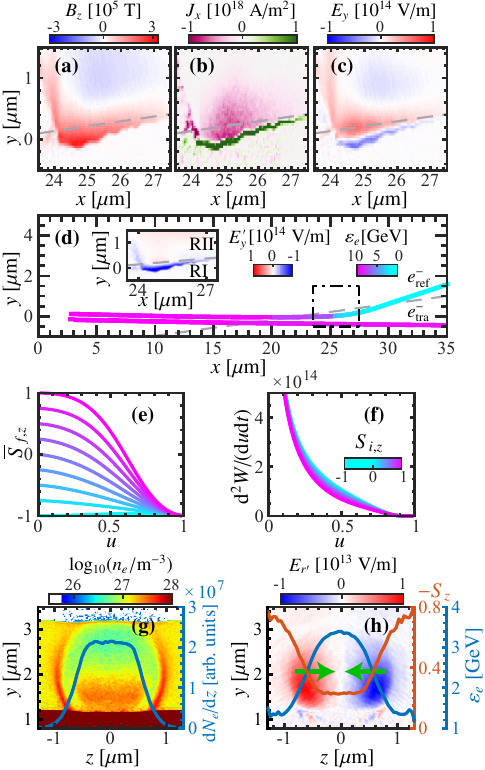}	
	\caption{ Spatial Distributions of (a) the magnetic field $B_z$, (b) current $J_x$ and (c) electric field $E_y$ in the $xy$ plane at $t=22T_0$.
		(d) Typical trajectory of the reflected and transmitted electrons, color coded by the electron energy $\varepsilon_e$. The inset displays the effective field $E^\prime_y$ at $z=0$ within the dash-dotted box at $t=22T_0$. The gray-dashed lines in (a)-(d) represent the initial interface between two target layers.  (e) Theoretical values of average polarization $\overline{S}_{f,z}$ and (f) probabilities d$^2W$/(d$u$d$t$) for various electron polarizations before radiation $S_{i,z}$, where $\chi_e=2$, $\varepsilon_e=10$~GeV, $\hat{\mathbf{v}}=(1,0,0)$ and $\hat{\mathbf{a}}=(0,1,0)$ are employed. Spatial distributions  of (g) the total electron density $n_e$ and (h) the focusing field $E_{r'}$ in the plane with $x=37.5$~$\mu$m at $t=35T_0$. The blue line in (g), red and blue lines in (h) represent the number distribution d$N_e$/d$z$,  polarization degree $S_z$, and energy $\varepsilon_{e}$ of the reflected electrons at this location, respectively. The green arrows in (h) represent direction of the focusing force on the reflected electrons.}
	\label{fig3}
\end{figure}

\begin{figure}[t]	
	\centering
	\includegraphics[width=1.0\linewidth]{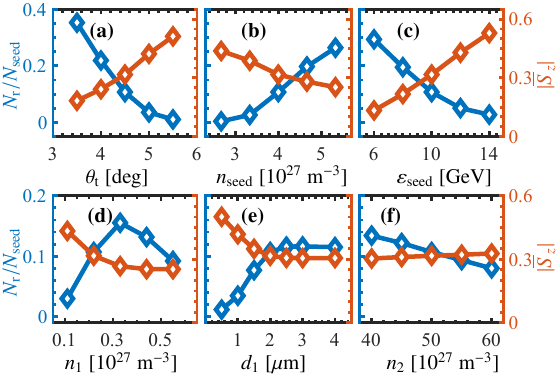}	
	\caption{Conversion rate $N_{\rm r}$/$N_{\rm seed}$ (blue) and average polarization $|{\overline{S}_z}|$ (red) of the reflected electron beam vs (a) the target tilt angle $\theta_{\rm t}$, (b) the seed beam density $n_{\rm seed}$, (c) the seed beam energy $\varepsilon_{\rm seed}$, (d) density of the first target layer $n_{1}$,  (e) depth of the first target layer $d_{1}$, and (f) density of the second target layer $n_{2}$.}
	\label{fig4}
\end{figure}

In this Letter, we put forward a novel method for the construction of a compact efficient ``polarizer'' that achieves direct ultrafast conversion of relativistic electron beams into polarized ones, based on the beam ``self-polarization'' mechanism via simple beam-target interactions; see the interaction scenario with a double-layer solid target, as an example, in Fig~\ref{fig1}(a). 
Specifically, when an unpolarized seed electron beam grazes through a solid target, it ionizes the target and excites an asymmetric plasma field due to the plasma electron backflows. This field then reacts on the beam itself, triggering spontaneous radiative polarization and reflection of the beam, and ultimately yielding a dense polarized electron beam; see Fig.~\ref{fig1}(b). We also find that compared to a single-layer target setup, the double-layer target setup can induce a bubble-like density perturbation region that further focuses the reflected electrons and reshapes its polarization distribution [see the schematic in Fig.~\ref{fig1}(c) and the simulation results in Figs.~\ref{fig2}(c) and (d)], which suggests that further improvement could be achieved by optimizing the target setup. The underlying mechanisms of beam self-polarization, focusing, and spin-reshaping are further detailed in Fig.~\ref{fig3}. Three-dimensional spin-resolved QED particle-in-cell (PIC) simulations demonstrate the production of a picocoulomb electron beam with a current of approximately 0.4~kA and a polarization degree of 65\%, indicating great potential for polarized beam experiments~\cite{moortgat2008polarized,flottmann1993,moortgat2008polarized,duan2019concepts, lin2018polarized}. Moreover, our method exhibits considerable robustness with respect to the beam and target parameters; see Fig.~\ref{fig4}. 

The spin-resolved QED-PIC code SLIPs~\cite{wan2023simulations} is employed to simulate the beam-plasma interactions, where the spin-resolved electron dynamics, stochastic photon emissions, and corresponding stochastic radiative spin evolution are described by a Monte Carlo algorithm. The NCS process is implemented, characterized by the nonlinear QED parameter $\chi_e\equiv|e|\hbar/(m_e^3c^4)\sqrt{-(F_{\mu\nu}p^{\nu})}$. Here, $F_{\mu\nu}$ is the electromagnetic field tensor, $p^\nu$ is the electron four-momentum, $\hbar$ is the reduced Planck constant, $e$ and $m_e$ are the electron charge and mass, respectively, and $c$ is the speed of light in vacuum. More details can be found in the Supplemental Material (SM)~\cite{SM}.

Typical results are shown in Fig.~\ref{fig2}. The simulation window has dimensions of $x\times y \times z=12~\mu$m $\times14~\mu$m $\times12~\mu$m, with grid cells of 600 $\times$ 700 $\times$ 600, moving at the speed $c$ along $+\hat{{x}}$. The seed electron beam moves along $+\hat{{x}}$, with 6~nC charge, 10~GeV mean energy,  5\% energy spread, 2$^{\circ}$ angle spread, and a Gaussian distribution of $n_{\rm seed}\exp\left(-{x^2}/{\sigma_x^2}-{r^2}/{\sigma_\perp^2}\right)$, where $n_{\rm seed}=4\times10^{27}$~m$^{-3}$, $r=\sqrt{y^2+z^2}$, $\sigma_x$ = 1.7~$\mu$m and $\sigma_\perp$ = 1.2~$\mu$m. Such electron beams are anticipated at FACET-{\uppercase\expandafter{\romannumeral2}} by employing a plasma lens or magnetic pinching~\cite{yakimenko2019FACET,yakimenko2019prospect,doss2019laser,zhu2023magnetic}, and may become available soon through advanced acceleration techniques~\cite{thevenet2016vacuum,shi2021generation,babjak2024direct}.  A double-layer carbon target is placed at an angle $\theta_{\mathrm{t}}=4.5^\circ$ to the $xz$-plane. Densities of the first and second layers are $n_1$ = $0.22\times10^{27}$~m$^{-3}$ and $n_2$ = $50\times10^{27}$~m$^{-3}$, respectively, and the layer thicknesses are $d_1$ = $d_2$ = 2~$\mu$m. Note that low-$Z$ target materials are required to effectively suppress bremsstrahlung and Bethe-Heitler process, as these process are negligible for high-energy electrons traversing a low-$Z$ thin target~\cite{heitler1954quantum}; see more details in SM~\cite{SM}. Each cell contains 16 macroparticles for the beam and target electrons, and 8 for the carbon ions C$^{6+}$. Note that the field driven by the seed beam is sufficient to fully ionize the carbon target in the primary interaction region, see more in SM~\cite{SM}.

Angular and polarization distributions of the reflected electrons are shown in Figs.~\ref{fig2}(a) and (b), respectively. These electrons have an average polarization degree exceeding 32\% and a charge of approximately 0.65~nC, with a conversion rate $N_{\rm r}/N_{\rm seed}$ $\approx$ 10.8\%. Here $N_{\rm r}$ and $N_{\rm seed}$ are numbers of the reflected and seed electrons, respectively. 
Their density is $\sim5\times10^{26}$~m$^{-3}$, which is three orders of magnitude higher than those in laser-electron beam collision schemes ($\sim$ 10$^{23}$~m$^{-3}$)~\cite{song2019spin,li2019ultra,seipt2019ultrafast}.  
At the final time $t=60T_0$ (with $T_0$ $\approx$ 3.33~fs), the electrons are mainly distributed around $\theta$ = 9$^\circ$, with a full width at half maximum (FWHM) of $\Delta\theta$ $\approx$  3$^\circ$; see Fig.~\ref{fig2}(c). As $\theta$ increases, the polarization degree $|S_z|$ initially remains around 18\% and begins to increase at $\theta$ $\approx$ 8.5$^\circ$ due to the dependence of the radiative polarization on energy of the emitted photons; see more details in Fig.\ref{fig3}(e). 
The number distribution peaks at $\phi$ = 0$^\circ$, with a FWHM of $\Delta\phi$ $\approx$ 30$^\circ$, and $|S_z|$ $\approx$ 43\% at $\phi$ = $0^\circ$; see Fig.~\ref{fig2}(d). 
Compared to results at $t=25T_0$ [see the dash-dotted lines in Figs.~\ref{fig2}(c) and (d)], the focusing effect reduces $\Delta\theta$ and $\Delta\phi$ by 40\% and 50\%, respectively, and increases $|S_z|$ at $\phi = 0^\circ$ by 38\%. Using a single-layer target can also achieve an average polarization of $\sim$29\% and a conversion rate of about 9\%. However, without focusing and spin-reshaping, the FWHM of the azimuthal angle $\Delta\phi$ is about twice that shown in Fig.~\ref{fig2}(d), and $|S_z|\approx$ 26\% at $\phi = 0^\circ$. See more details on using a single-layer target in SM~\cite{SM}.

To facilitate electron capture and transfer for subsequent applications, we focus on electrons within the peak cone angle $\alpha =$ ($|\theta-9^\circ|\leq 1^\circ, |\phi| \leq 1^\circ$). These electrons have an average polarization of $|\overline{S}_z|\approx$ 37\%, significantly higher than what would be achieved with asymmetric laser pulses ($\lesssim$ 10\%)~\cite{song2019spin,seipt2019ultrafast}.  
Their polarization $|S_z|$ decreases from 75\% to 16\% within the FWHM energy range of 1.0~GeV $<\varepsilon_{e}<$ 3.1~GeV, and $|S_z|\approx$ 65\% at the energy peak of $\varepsilon_e\approx1.5$~GeV; see Fig.~\ref{fig2}(e). These electrons have a charge of about 21~pC and an angle divergence of $3.5\times3.5$~mrad$^2$, with transverse and longitudinal sizes of $0.5\times3.0$~$\mu$m$^2$ and 1.0~$\mu\mathrm{m}$, respectively, at $t=60T_0$. The electric current can reach $\sim6.3$~kA, comparable to that of polarized electron beams from plasma wakefield acceleration~\cite{wen2019polarized,wu2019polarizedNJP,nie2021in,nie2022highly,bohlen2023colliding,gong2023spin,sun2024generation}, and significantly exceeding those achieved in laser-electron beam collision schemes ($\sim$ 10~A)~\cite{li2019ultra,song2019spin,seipt2019ultrafast} and traditional methods (0.1~A $-$ 1~A)~\cite{mane2005spin}.
Through further energy selection, electron beams with various polarization degree can be obtained. For example, the polarization degrees are $S_z$ $\approx$ 75\%, 65\%, 49\%, and 19\% for $\varepsilon_e$ = 1.0, 1.5, 2.0, and 3.0 GeV, respectively. The corresponding brilliances are 0.49$\times10^{25}$, 1.4$\times10^{25}$, 1.6$\times10^{25}$, and 1.5$\times10^{25}$~$e$/(s mm$^2$ mrad$^2\times$ 0.1\% bandwidth), respectively. Moreover, at the energy peak of $\varepsilon_e$ = 1.5~GeV, with a 10\% energy spread, the electrons have a charge of 1.3~pC and a current of 0.39~kA. Such polarized beams, with exceptionally high current ($\sim$ kiloampere) and polarization ($\gtrsim$ 60\%), holds great promise for related strong-field QED research~\cite{yousef2006relativistic,piazza2012extremely,gonoskov2022charged,fedotov2023advances}. The reflected electrons rapidly pass through the weak focusing field and, therefore, experience very little depolarization~\cite{thomas2020scaling}; see Fig.~\ref{fig2}(f).

Figure~\ref{fig3} illustrates the main factors responsible for self-polarization, namely focusing and spin-reshaping. As the seed electrons graze through the polarizer, an ultrastrong asymmetric magnetic field is excited between the two target layers due to the plasma electron backflow, equivalent to the current density $\mathbf{J}_{b}$ = $-en_{b}\mathbf{v}_{b}$, for charge balance~\cite{wu2009generation,cai2011magnetic}; see Figs.~\ref{fig3}(a) and (b). This follows approximately from $\nabla\times\mathbf{B} = \mu \mathbf{J}_{b}$, with a maximum $|\mathbf{B}|$ $\approx$ $|e|n_{\rm seed}\sigma_\perp/(\varepsilon_0c)$ $\sim$ 10$^5$~T~\cite{zhu2023magnetic}. Here, $n_{b}$ and $\mathbf{v}_{b}$ are the density and velocity, respectively, of the plasma backflow, and $\varepsilon_0$ is the vacuum permittivity.
Using a double-layer target can increase density of the seed beam via self-focusing in the plasma~\cite{cox1975self,mccorkle1975ele}, as the seed beam passes through the first target layer, thereby enhancing the magnetic field; see SM~\cite{SM}. Also, the seed beam rapidly pushes the plasma electrons away, while the massive ions remain nearly stationary, creating a space charge field $\mathbf{E}$; see Fig.~\ref{fig3}(c).

In these fields, due to stochastic nature of the quantum radiation~\cite{shen1972energy,duclous2011monte,blackburn2014quantum,harvey2017quantum,li2018electron,li2014robust,li2017angle,wan2019imprint,wan2022quasi}, electrons that experience a large energy loss $\Delta\varepsilon_{e}$ are reflected, while the rest continue through the target. This can be explained by a simplified model.
Since the seed electrons propagate mainly along $+\hat{{x}}$, the space can be divided into two regions, based on the effective field $\mathbf{E}'$ = $\mathbf{E}_\perp+\mathbf{v}\times\mathbf{B}$ experienced by the electrons: in Region I (RI), $\mathbf{E}'$ $\parallel$ $-\hat{{y}}$ with $|\mathbf{E}'|$ reaching about 10$^{14}$~V/m, while in Region II (RII), $\mathbf{E}'$ $\parallel$ $\hat{{y}}$ with a maximum $|\mathbf{E}'|\sim10^{13}$~V/m; see Fig.~\ref{fig3}(d).
Here $\mathbf{E}_\perp$ is the electric field component perpendicular to the electron velocity vector $\mathbf{v}$.
The electron deflection angle is given by $\theta_y = \arctan(p_y/p_x)$, with $p_i\approx p_{i0}-e\int E'_{i}\mathrm{d}t-\int p_iP_{\rm rad}/(\gamma_e m_ec^2){\rm d}t$, where $i=x$ or $y$, $p_{i0}$ is the initial $i$-momentum, $P_{\rm rad}$ is the radiation power, $\gamma_e$ is the electron Lorentz factor, and the third term represents radiation reaction. 
In RI, the nonlinear QED parameter $\chi_e = \gamma_e|\mathbf{E}'|/E_{\rm c}\sim$ 2, leading to significant photon emission. Here, $E_c = m_e^2c^3/(|e|\hbar)\approx 1.3 \times 10^{18}$~V/m is the QED critical field strength~\cite{schwinger1951gauge}. Meanwhile, the Lorentz force $\mathbf{F} = e\mathbf{E}'\parallel+\hat{{y}}$ deflects electrons upwards ($+\hat{{y}}$ direction). 
Thus, electrons with a large energy loss are reflected into RII, where the weak Lorentz force $\mathbf{F}\parallel-\hat{{y}}$ cannot push them back into RI. Ultimately, these electrons move upward and leave the target, i.e., get reflected; see a typical reflected electron trajectory in Fig.~\ref{fig3}(d). By contrast, electrons with a small energy loss retain enough $x$-momentum to pass through RI, i.e., are transmitted.
Moreover, the reflected electrons originate mainly from the rear of the seed beam, since the plasma field is relatively weak when the front electrons pass through the target; see SM~\cite{SM}.

Average polarization of an electron after emitting a photon is given by~\cite{baier1998,SM} $\overline{\mathbf{S}}_f = \mathbf{g}/\left[w + u K_{1/3}(\rho)\hat{\mathbf{b}}\cdot \mathbf{S}_i\right]$, where $\mathbf{g} = \frac{-u}{1-u}K_{1/3}(\rho)\hat{\mathbf{b}} + \left[2K_{2/3}(\rho) - {\rm Int}K_{1/3}(\rho)\right]\mathbf{S}_{i} + \frac{u^2}{1-u}\left[K_{2/3}(\rho) - \mathrm{Int}K_{1/3}(\rho)\right](\mathbf{S}_i\cdot\hat{\mathbf{v}})\hat{\mathbf{v}}$, $w = \frac{u^2-2u+2}{1-u}K_{2/3}(\rho) - \mathrm{Int}K_{1/3}(\rho)$, $u = \varepsilon_\gamma/\varepsilon_{i}$, $\varepsilon_{i}$ and $\mathbf{S}_i$ are the electron energy and spin, respectively, before radiation, $\varepsilon_\gamma$ is the photon energy, $K_n$ is a modified Bessel function of the second kind of order $n$, $\mathrm{Int}K_{1/3}(\rho)=\int_{\rho}^{\infty} {\rm d}z {K}_{\frac{1}{3}}(z)$, $\rho = 2u/[3(1-u)\chi_e]$, $\hat{\mathbf{b}} = \hat{\mathbf{v}}\times\hat{\mathbf{a}}/|\hat{\mathbf{v}}\times\hat{\mathbf{a}}|\approx\mathbf{B}'/|\mathbf{B}'|$, $\mathbf{B}'\approx\gamma_e[\mathbf{B}-\hat{\mathbf{v}}\times\mathbf{E}-\hat{\mathbf{v}}(\hat{\mathbf{v}}\cdot\mathbf{B})]$ is the magnetic field in rest frame of the electron, and $\hat{\mathbf{v}}$ and $\hat{\mathbf{a}}$ are unit vectors along the electron velocity and acceleration, respectively. As a result, the electron spin after radiation tends to align along ${\bm \zeta}$ = $-\mathbf{B}'/|\mathbf{B}'|$.
In the ultrarelativistic limit ($\gamma_e\gg1)$, $\mathbf{B}'/\gamma_e\approx(0, 0, B_z-v_xE_y/c)\approx(0,0,-E'_y/c)\parallel +\hat{z}$ in RI, while $\mathbf{B}'/\gamma_e\parallel -\hat{z}$ in RII. Since most radiation occurs in RI, electrons can build up radiative polarization along ${\bm \zeta}\parallel-\hat{z}$.
Although low-energy photons are more likely to be emitted, their impact on electron spin is small (i.e., as $u\rightarrow0, \overline{S}_{f,z}\rightarrow S_{i,z})$; see Figs.~\ref{fig3}(e) and (f). Conversely, high-energy photon emission significantly aligns the spin with ${\bm \zeta}$ (i.e., as $u\rightarrow1, \overline{S}_{f,z}\rightarrow-1)$, resulting in high polarization for low-energy electrons; see Fig.~\ref{fig2}(e). The selection effect during non-radiative periods further influences the electron spin dynamics~\cite{cain}, leading to polarization of the high-energy electrons along $+\hat{{z}}$; see more details in SM~\cite{SM}.  
Electrons with large polar angles generally experience great energy loss, leading to high polarization; see Fig.~\ref{fig2}(c).

Focusing and spin-reshaping arise from plasma bubbles in the first target layer, where the seed beam pushes plasma electrons away, leaving ions behind; see Fig.~\ref{fig1}(b) and Fig.~\ref{fig3}(g). As the reflected electrons pass through the bubble, their angle spread is effectively decreased by a radial focusing field $E_{r'} \approx (k_pr'/2)E_0\sim$ 10$^{12}$~V/m~\cite{esarey2009physics}; see Fig.~\ref{fig3}(h). Here $r'$ is perpendicular distance of the electron from the axis of $E_{r'}, k_p = \omega_p/c, E_0 = cm_e\omega_p/|e|$, and $\omega_p = \sqrt{n_ee^2/(\varepsilon_0m_e)}$ is the electron plasma frequency. 
Electrons with larger azimuthal angles $\phi$ experience stronger focusing forces, since $E_{r'}\propto r'$, and acquire higher polarization due to significant energy loss; see the red and blue lines in Fig.~\ref{fig3}(h). 
This results in significant reshaping of both $\phi$-resolved number and polarization distributions of the reflected electrons; see Fig.~\ref{fig2}(d). Moreover, as particle acceleration and beam control techniques advance~\cite{esarey2009physics,hooker2013developments,donald2003relativistic,bingham2004plasma,agostini2021large,gschwendtner2022awake,caldwell2016path,sha2022dense}, especially in increasing density, our method may also be applied to other relativistic dense charged particle beams, which can similarly drive asymmetric plasma fields.

To demonstrate experimental feasibility, impact of the seed beam and target parameters on the conversion rate $N_{\rm r}/N_{\rm seed}$ and the average polarization degree $|\overline{S}_z|$ of the reflected electrons, is investigated in Fig.~\ref{fig4}. First, the target tilt angle $\theta_{\rm t}$ plays a crucial role.  As $\theta_{\rm t}$ decreases, electrons gradually begin to oscillate between RI and RII (see SM~\cite{SM}), enhancing radiation and increasing $N_{\rm r}/N_{\rm seed}$; see Fig.~\ref{fig4}(a). However, $|\overline{S}_z|$ decreases due to enhanced radiation in RII, where polarization builds oppositely. Conversely, a large $\theta_{\rm t}$ leads to reduced radiation and decrease $N_{\rm r}/N_{\rm seed}$. Thus, an appropriate $\theta_{\rm t}$ can be chosen to meet different requirements for electron polarization and charge. Increasing density of the seed beam $n_{\rm seed}$ strengthens the effective field $|\mathbf{E}'|\propto n_{\rm seed}$ and electron energy loss $\Delta \varepsilon_e\propto\chi_e\propto|\mathbf{E}'|$~\cite{piazza2012extremely}, leading to an increase in $N_{\rm r}/N_{\rm seed}$; see Fig.~\ref{fig4}(b).  However, $|\overline{S}_z|$ decreases due to enhanced radiation in RII. For electron beams with a low density, optimizing the target setup or applying additional focusing techniques, such as magnetic pinching~\cite{zhu2023magnetic}, can enhance $|\mathbf{E}'|$ and increase $N_{\rm r}/N_{\rm seed}$.  A higher energy of the seed beam $\varepsilon_{\rm seed}$ decreases $N_{\rm r}/N_{\rm seed}$, due to greater initial $x$-momentum but increases $|\overline{S}_z|$ as electrons must undergo more energy loss before being reflected; see Fig.~\ref{fig4}(c).
As density of the first target layer $n_1$ increases, self-focusing effect of the seed beam strengthens, resulting in a stronger effective field. Consequently, similar to the effects of increasing $n_{\rm seed}, N_{\rm r}/N_{\rm seed}$ initially increases while $|\overline{S}_z|$ decreases; see Fig.~\ref{fig4}(d). However, beyond a certain $n_1$ ($\approx$ $0.3\times10^{27}~m^{-3}$ for the given parameters), the effective field strength in RII is sufficient to give the seed beam a non-negligible negative $y$-momentum before it enters RI. This causes effects similar to those of increasing $\theta_{\rm t}$, namely a decrease in $N_{\rm r}/N_{\rm seed}$. Similarly, increasing depth of the first target layer $d_1$ initially raises $N_{\rm r}/N_{\rm seed}$ and decreases $|\overline{S}_z|$ until saturation is reached at $d_1\approx2\sigma_\perp = 2.4~\mu$m; see Fig.~\ref{fig4}(e). Finally, as density of the second target layer $n_2$ increases, spatial scale of the effective field in RI, characterized by the plasma skin depth $l_s = c/\omega_{p}\propto\sqrt{1/n_2}$, decreases, reducing the $x$-momentum required for electron transmission, and thereby decreasing $N_{\rm r}/N_{\rm seed}$; see Fig.~\ref{fig4}(f). Meanwhile, $|\overline{S}_z|$ slightly increases due to a lower reflected electron energy $\varepsilon_{e}\propto|p_x|\propto-|\Delta p_x|$. Therefore, our method exhibits considerable robustness with respect to the beam and target parameters.

In conclusion, we have put forward a compact efficient polarizer for relativistic dense electron beams, based on the beam self-polarization mechanism via simple beam-target interactions. Our method opens up a new avenue for relativistic beam polarization with compact accessible devices, which would enhance their broad range of applications and facilitate development of related experiments, such as in strong-field QED studies~\cite{yakimenko2019prospect,piazza2012extremely,gonoskov2022charged,fedotov2023advances}, and polarized electron-positron and electron-ion colliders~\cite{flottmann1993,moortgat2008polarized,duan2019concepts, lin2018polarized}.

\hspace*{\fill}

\textit{\textbf{Acknowledgement}} 
This work is supported by the National Natural Science Foundation of China (Grant Nos. 12425510, 12275209, 12375244,  12135009, and U2267204), the Foundation of Science and Technology on Plasma Physics Laboratory (Grant No. JCKYS2021212008), the Fundamental Research Funds for Central Universities (Grant No. xyz012023046), the Natural Science Basic Research Program of Shaanxi (Grant No. 2023-JC-QN-0091), the Shaanxi Fundamental Science Research Project for Mathematics and Physics (Grant Nos. 22JSY014, and 22JSQ019), and the Hunan Provincial Research and Innovation Foundation for Graduate Students (Grant No. CX20220048). YIS is supported by an American University of Sharjah Faculty Research Grant (FRG24-E-S29).

\bibliography{mybib}

\end{document}